\documentclass[11pt,a4paper,review,3p,times,authoryear]{elsarticle}

\usepackage{graphicx}
\usepackage{amssymb}
\usepackage{amsmath}
\usepackage{mathtools}
\usepackage{dsfont}
\usepackage{color}
\usepackage{url}
\usepackage{natbib}

\usepackage{dirtytalk}
\usepackage{amsthm}
\newtheorem{definition}{Definition}

\usepackage{lineno}

\journal{Decision Support Systems}

\begin{document}

\begin{frontmatter}

%xxx improve title....?
\title{High-performance stock index trading: making effective use of  a deep long short-term memory network}



\author[rvt]{Chariton Chalvatzis\corref{cor1}}
\ead{charis.chalvatzis@uom.edu.gr}
\cortext[cor1]{Corresponding author}
\address[rvt]{University of Macedonia, Department of Applied Informatics, Egnatia 156, Thessaloniki, 54006, Greece}

\author[dcv]{Dimitrios Hristu-Varsakelis\fnref{rvt}}
\ead{dcv@uom.gr}

\begin{abstract}
We present a deep long short-term memory (LSTM)-based neural network for predicting asset prices, together with a successful trading strategy for generating profits based on the model's predictions. 
Our work is motivated by the fact that the effectiveness of any prediction model is inherently coupled to the trading strategy it is used with, and vise versa. This highlights the difficulty in developing models and strategies which are jointly optimal, but also points to avenues of investigation which are broader than prevailing approaches.
Our LSTM model is structurally simple and generates predictions based on price observations over a modest number of past trading days. The model's architecture is tuned to promote profitability, as opposed to accuracy, under a strategy that does not trade simply based on whether the price is predicted to rise or fall, but rather takes advantage of the  distribution of predicted returns, and the fact that a prediction's position within that distribution carries useful information about the expected profitability of a trade.
The proposed model and trading strategy were tested on the S$\&$P $500$, Dow Jones Industrial Average (DJIA), NASDAQ and Russel $2000$ stock indices, and achieved cumulative returns of $340\%$, $185\%$, $371\%$ and $360\%$, respectively, over $2010$-$2018$, far outperforming the benchmark buy-and-hold strategy as well as other recent efforts.
\end{abstract}


\end{frontmatter}

%\linenumbers
\linespread{1.3}

\section{Introduction}
\label{S:1}

Equity prediction lies at the core of the investment management profession and has also attracted significant attention from academia. One of the open questions has been whether (and how) one can forecast the behavior of stocks and then act accordingly to generate ``excess returns'', i.e., profit in excess of those generated by the market itself \citep{Fama_French}.
Towards that end, significant effort has been put into predicting the price of major U.S. stock indices such as the S\&P~500 or the Dow Jones Industrial Average (DJIA)  \citep{Huck_2009, Huck_2010, Sethi_Treleaven_Rollin, krauss_do_huck, Bao} as well as that of individual stocks, using techniques ranging from early linear models \citep{fama_french_CAPM} 
to machine learning and neural network-based approaches \citep{Sermpinis_EJOR_SwarmOpt,deep_RL, dl_GRU}.
There are two components which are generally part of the overall discussion on ``intelligent'' or automatic trading: a {\em predictive model} whose purpose is to anticipate an asset's future price, and a {\em trading strategy} which uses the model's predictions to generate profit. Developing a scheme with consistently superior performance is difficult, in part because of the ``joint'' nature of the goal: profits ultimately depend both on the prediction model as well as the trading strategy it is being used with, and changing either of the two affects the final outcome. Thus, in principle, these two components should be designed and optimized together, a challenge which currently appears to be out of reach, due to the seemingly endless variety of possible models and trading strategies. Instead, most of the relevant research (to be reviewed shortly) has focused on schemes that do quite well by improving on specific aspects of the overall prediction/trading process. 

One approach has been to aim for a prediction model which is ``as good as possible'' at guessing the ``next'' price of an asset, and then use that guess to inform trading decisions \citep{Bao, ModAugNet, EMD2FNN}. This is frequently done by means of a so-called {\em directional} \say{up-down} trading strategy which buys the asset when the predicted price is higher than the current price and sells if it is lower. While this approach can be effective, it 
is based on an implicit assumption that better short-term price predictions lead to higher profitability. This is not true in general: prediction accuracy is typically measured with a {\em symmetric} loss function (e.g., mean absolute error) which is indifferent to the sign of the prediction error, and yet that sign may be quite important when trading. In fact, it is possible for one model to lead to trades which are {\em always} profitable while another, having {\em better} predictive accuracy, to record only losses\footnote{Consider for example the four-day price series $Y=[101, 100, 98, 101]$, and two possible predictions of its last three samples, $P_1 = [95, 92, 105]$ and $P_2 = [102, 101, 97]$, both generated on the first day, when the price was 101. $P_1$ has a mean absolute error of 5 when measured against the last three samples of $Y$, while $P_2$'s error is 3. However, $P_1$ is always correct on the direction of the price movement of $Y$, i.e., its predictions are always higher (resp.lower) than the previous price of $Y$ when the price will indeed rise (resp. fall) in the next sample, while $P_2$ is always wrong.}. 
An alternative is to forgo predictions of the asset price itself and instead consider a {\em directional} model which predicts, with as high an accuracy as possible, whether the price will rise or fall compared to its current level, essentially acting as a binary classifier \citep{KraussFisher,Zhong_Enke,adaptive_stock_index}. Then, the trading strategy is again to simply buy or sell the asset based on the model's ``recommendation''. This approach is ``self-consistent'' from the point of view that the model's predictions are judged against in the same setting as the trading strategy, i.e., being correct on the direction of price movement. 

While one can of course consider many possible variations on trading strategies and accompanying models, there are two key points to be recognized. First, prediction models should be viewed and evaluated  {\em in the context of the trading strategies} they are used with \citep{Leith_Tanner_Forecast}. Thus, prediction accuracy should not monopolize our focus: large(r) prediction errors do not necessarily preclude a model from being profitable under the ``right'' trading strategy, just like low errors do not by themselves guarantee profitability. Second, because the joint model/strategy optimum appears to be elusive, one may try to  
improve overall profitability by {\em adapting the trading strategy} to 
to make ``fuller'' use of the information contained within the model's predictions; that information may generally be much more than whether the price will rise or fall in the next time step.

This paper's contribution is to build on the ideas outlined in the previous paragraph by proposing a novel model-strategy pair which 
together are effective in generating profits with reasonable trading risk, and compare favorably to standard benchmarks as well as recent works.
Our model  consists of a deep long short-term memory (LSTM)-based neural network which is to predict an asset's price based on a rolling window of past price data. Structurally, the network will be simple so that it can be trained and updated fast, making it suitable for intra-day trading, if desired. A key feature will be that the network's output layer will have access to the entire evolution of the network's hidden states as the input layer is ``exposed'' to a sequence of past prices. This will allow us to achieve a predictive accuracy similar to that of more complex models, while keeping our architecture ``small'' in terms of structure and data required.
%using only a small number of layers ($\le 3$) and a few data features related to an asset's daily price over a modest number of past trading days (22 or fewer). 
Our LSTM network will initially be trained to achieve low price prediction errors, as opposed to high profitability which is our ultimate goal. We will bring profitability into the fold 
in two ways: a) the network's hyper-parameters will be selected to maximize {\em profit} (instead of accuracy, as is typically done in the literature 
\citep{Zhong_Enke,
%use trial and error for hyper while 
EMD2FNN}) 
%use grid search and MSE.  
in conjunction with  b) a novel event-based\footnote{We will use the term ``event'' in the sense of probability theory (e.g., a sample of a random variable falling within a specified interval on the real line), as opposed to finance where it commonly refers to ``external'' events (corporate filings, legal actions, etc.).} 
trading strategy which seeks to take advantage of the information available not in any single price prediction, but in the entire {\em distribution} of predicted returns.   
In broad terms, our strategy will 
not  act simply on whether the predicted price is higher or lower compared to its current value, but will instead make decisions based on the expected profitability of a possible trade, conditioned on the prediction's relative position within the aforementioned distribution. As we will see, such a strategy can be optimized in some ways and will lead to significant gains. 

We will test our proposed approach on major U.S. stock indices, namely the S\&P 500 (including dividends), the Dow Jones Industrial Average (DJIA), the NASDAQ and the Russel 2000 (R$2000$), over an almost 8-year long period (October 2010 - May 2018). These indices attract significant attention from market practitioners, both in terms of strategy benchmarking and for monitoring market performance. They will also serve as a common ground for comparison with other studies. 
When it comes to discussing the performance of our approach, we will provide the ``standard'' annualized and cumulative profits attained, but also additional metrics that relate to trading risk, 
which are very much of interest when assessing any investment but are often missing from similar studies. Moreover, our relatively long testing period  will allow us to draw safer conclusions as to our scheme's profitability over time.

The remainder of the paper is organized as follows. Section~\ref{sec:litreview} discusses relevant literature, including details on profitability and overall performance. In Section~\ref{sec:model} we present our proposed prediction model and trading strategy, as well as a class of capital allocation policies which are optimal for our trading strategy. Section~\ref{sec:training} discusses model training and allocation policy selection. The profitability and overall performance of our model and strategy, as well as comparisons with recent works and ``naive'' benchmark approaches of interest, are discussed in Section \ref{sec:results}. 

\section{Literature Review}
\label{sec:litreview}

We can identify two significant clusters in the literature which are most relevant to this work. One contains studies in which prediction and profitability evaluation are ``disconnected'' in the sense discussed in the previous Section: a model is trained to predict the value of an asset, but that prediction is then used with a \say{directional} trading strategy. 
Studies that fall within this category include \citet{Bao} who described a three-step process (de-noising of the stock price series, stacked auto-encoders, and a recurrent neural network using LSTM cells) achieving strong results, including a sum of daily returns whose yearly average was approximately $46\%$ and $65\%$ for the S$\&$P $500$ and DJIA, respectively, over the period $2010$-$2016$.
\citet{ModAugNet} explored the role of limited data when training neural networks and the resulting deterioration in prediction accuracy, and introduced a modular architecture consisting of two LSTM networks. Using a 
$20$ trading day window and the S$\&$P$500$ as their underlying asset
the authors reported low prediction errors (e.g., a 12.058 mean absolute error)
%($1.0759$ mean absolute percent error, $342.48$ mean squared error and $12.058$ mean absolute error) 
and a $96.55\%$ cumulative return\footnote{Cumulative return (CR) is the product of the sequence of gross returns over a given time period, minus unity, i.e., $CR=\prod_2^N s_i/s_{i-1}-1$ for a stock or other asset whose value is $s_i$ over trading days $i=1,2,\ldots,N$.}, beating the benchmark buy-and-hold\footnote{A standard benchmark for gauging the performance of a trading strategy over a given time period, is the so-called {\em buy-and-hold} 
(BnH) 
portfolio which is formed simply by buying the asset on the first day of the period under consideration, and selling it on the last day.} strategy ($78.74\%$) over the period from $2/1/2008$ to $7/26/2017$ (we will conventionally use the U.S. format, $mm/dd/yyyy$ for reporting dates throughout).

Several studies have applied feed-forward or convolutional neural networks (CNN) to financial time-series. A recent example is \citet{EMD2FNN}, which introduced the notion of empirical mode decomposition and a factorization machine-based neural network to predict the future stock price trend. 
%Applied to the S$\&$P$500$ (over $1/3/2007$-$12/30/2011$), and the NASDAQ and Shanghai Stock Exchange Composite (SSEC) indices ($1/4/2012$-$12/30/2016$). 
For the S$\&$P$500$, using a one-year test period (the year $2011$), the authors reported $13.039$, $17.6591$, and $1.05\%$ for the mean absolute, root mean squared and mean percentage errors, respectively. Combining their predictions with a long-short\footnote{ A Long-Short trading strategy is one in which in addition to buying, one is also allowed to sell short (borrowd) assets.} trading strategy on the same index, they reported an average annualized return\footnote{Annualized return (AR) is the per-year geometric average of returns over a given time period,  minus unity, i.e., $AR=(1+CR)^{252/N}-1$, for an asset whose value is $s_i$ over trading days $i=1,2,\ldots,N$, where 252 is the nominal number of trading days in a calendar year.} of $25\%$ and a Sharpe ratio\footnote{The Sharpe ratio (SR) is a measure of reward-to-risk performance, defined as the difference between AR and the federal funds rate (which is considered to be the \say{risk-free} rate, essentially at zero following the financial crisis of $2007-2008$), divided by the annualized volatility (AV) of the returns. AV is the standard deviation of returns over a given time period, multiplied by the square root of the number of trading days in a year (conventionally $252$).}  of $2.6$. 
Another interesting CNN-based approach is that of \citet{OmerMurat_Algorithmic} in which price time-series are transformed into $2$-D images which are then fed into a trading model to produce a ``buy'', ``sell'' or ``hold'' recommendation. Trading on well-known exchange-traded funds (ETFs) as well as the constituents of the DJIA, that approach could yield an annualized return of $13.01\%$ for the ETF portfolio and $12.59\%$ for the DJIA constituents portfolio, over  $2007$-$2017$. 
Finally, \citet{krauss_do_huck} is an example of using  a portfolio of stocks to outperform an index, as opposed to trading the index itself. That work tested several machine learning models, including deep neural networks, gradient-boosted trees and random forests in order to perform statistical arbitrage on the S\&P$500$ using daily predictions from $1992$ to $2015$. Before transaction costs, the authors 
%proposed an equal-weighted ensemble of models and, for the purposes of trading, formed a long-short portfolio of stocks instead of trading the index only; 
reported mean daily excess returns of $0.45\%$ for their ensemble model during $12/1992$-$10/2015$, however returns were negative during that period's last five years ($2010$-$2015$). 
%As the authors state, differences in the economic environment between the decades could be one reason for the model's ``instability''. 

%%%%%%%%%%%%%%%%%%%%%%%%%%%%%%%%%%%%%%%

A second important cluster of works contains studies in which the loss function used to train the prediction model is ``consistent'' with the trading strategy, e.g., the model predicts the {\em direction} of the price movement - but not the numerical value of the price - which is then  used with a directional trading strategy. 
%Trying to guess correctly - roughly speaking - as to whether an asset's closing price will rise or fall, essentially turns the problem into one of classification. 
\citet{Sethi_Treleaven_Rollin} presented a neural network which used only two features to predict the direction (up or down) of next-day price movement for the largest $100$ stocks (in terms of market capitalization) in the S\&P $500$ index. Those stocks were then combined into  a portfolio  in order to \say{beat} the index itself. Using data from $2006$-$2013$ and assuming a cost of $0.02\%$ per transaction, that approach achieved a $16.8\%$ annualized return. Recently, \citet{KraussFisher} also studied the daily direction of the S$\&$P $500$ constituents between $1/1/1990-9/30/2015$, comparing various machine learning techniques. 
%neural networks (using LSTM cells) against other well-known machine learning techniques (such as random forests) and on the other hand, analyze the sources of the profitability generated by the LSTM model. 
Their approach was to use a single feature, namely the standardized one day return over the past trading year, in order to predict the direction of the constituent stocks, and ultimately the probability for each stock to out-/under-perform the cross-sectional median.
%for the next period and finally, according to the ranking of the probabilities, they formed a long-short portfolio. 
The authors reported mean daily returns of $0.46\%$ and a Sharpe Ratio of $5.8$, which indicates low risk / high reward prior to transaction costs. However, the returns were not consistent throughout the testing period, and were close to zero during its last 5 years.
\citet{adaptive_stock_index} proposed a neural network for forecasting stock price movement, using a binary output to indicate buy/hold or sell trading suggestions.
%The authors propose a framework where each underlying asset has been trained individually and carries its own parameters, hence the name adaptive. 
The authors applied a denoising process to the input time-series before feeding it to the neural network. Their model, applied to a variety of ETFs over a 1-year period $1/1/2010$-$12/31/2010$, achieved a strong cumulative return of $41.89\%$ in the case of SPY\footnote{SPY is an Exchange Traded Fund (ETF) which tracks the S\&P500 index.}, before transaction costs, and even higher returns when short-selling was allowed. That approach also provides a good example of how very few trades ($14$ in one year) can yield good performance 
%($41.86\%$ after transaction costs) 
when trading individual assets.

The work in \citep{Zhong_Enke} also studied the daily direction of the S$\&$P $500$ index using  $60$ input features and various dimensionality reduction techniques. The study period covered $6/1/2003$-$5/31/2013$. 
The best model in that study reached a $58.1\%$ accuracy, however the accuracy of a ``naive'' or other baseline strategy over the same period was not reported. Using a trading strategy which either bought the S$\&$P $500$ or invested in a one-month T-bill
%, over a period of $376$ days. The best model 
resulted in a mean daily return of $0.1\%$ with a standard deviation of $0.7\%$ over their test period of  $377$ days ending on $5/31/2013$.
Finally, an interesting attempt at event-driven stock market prediction using CNNs is \citet{event-driven}. The model proposed therein also produces a binary output, (i.e., the stock price is predicted to increase or decrease). Using a directional trading strategy over a test period from $02/22/2013$-$11/21/2013$, that approach was applied to 15 stocks in the S$\&$P$500$, with a reported {\em mean} cumulative return of $167.85\%$.

It is important to note that several of the works cited here report {\em average returns} \citep{Bao, Zhong_Enke} but no annualized or cumulative returns. Unfortunately, the arithmetic average of returns is not informative as to actual capital growth / profitability, because it cannot be used to infer annual or cumulative return\footnote{For example, if we invest 1 euro in an asset which achieves consecutive returns of -99\%, 100\%, and 100\%, our investment would be worth just 4 cents, i.e., a cumulative return of -96\%, while the average return would be approximately 33\%.}.
Also, as one might expect, it is difficult to declare a single ``winner'' among the different approaches: not all authors report the same profitability metrics, and the evaluation periods vary; even when comparing against the same asset and testing period, one cannot safely draw conclusions if the testing period is rather short (e.g., a single year, as in \citet{ adaptive_stock_index, EMD2FNN, Zhong_Enke, event-driven}). 
Finally, there is also a significant number of works on the use of machine learning methods and neural networks in particular, where the goal is exclusively accurate price prediction, with no discussion of further use of that prediction, be it for trading or any other purpose. Representative papers include \citet{RNN_and_hybrid_model,chong_han_park_error_metrics, gu_Kelly_Xiu}. The volume of research in this category is substantial but we will not delve deeper here because prediction accuracy by itself is not the focus of this work.

\section{Proposed model and trading strategy}
\label{sec:model}
We proceed to describe a LSTM neural network which will be used to predict  stock or index prices, followed by a trading strategy that attempts to take advantage of the model's predictions - despite their inaccuracies - in a way which will be made precise shortly. 

\subsection{Network Architecture} \label{network_arch}

\begin{figure}[!ht]
\centering
\includegraphics[width=0.6\textwidth]{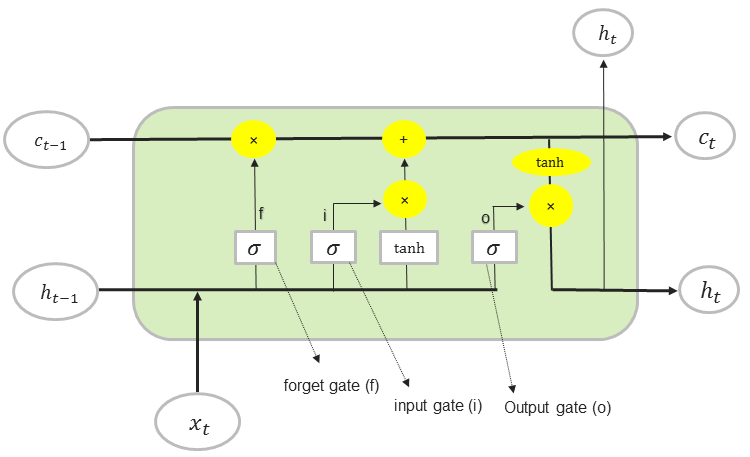}
\caption{LSTM cell diagram (based on  \citet{colahs}).}
\label{fig:LSTM}
\end{figure}

Our proposed network follows the popular LSTM architecture \citep{LSTM_origin}, commonly used in recurrent neural network applications \citep{image_recongnition_ILSVRC, AlphaGo, NMT, nmt_recent_advances}. 
%explain LSTM
The basic LSTM cell, shown in Fig.~\ref{fig:LSTM},
receives at each time $t=0,1,2,...$ a 
$B \times I$ batch of length-$I$ input vectors, $x_t$ (where $B$ is batch size and $I$ is the number of input features), 
and utilizes a number of multiplicative gates in order to control information flow through the network. In addition to the usual hidden state, $h_{t}$, of size $B \times H$, where $H$ is the number of hidden units, the LSTM cell includes an ``internal" or ``memory" state, $c_t$, also of size $B \times H$ in our case. 
%Both of these play a crucial role in the effectiveness of the network. 
The three gates, input ($i$), output ($o$) and forget ($f$), which have feed-forward and recurrent connections, receive the input $x_{t}$, 
and previous hidden state $h_{t-1}$, to produce
\begin{align}
  \label{eq_input1} 
    i_{t} &= \sigma(x_{t} W_{ix}  + h_{t-1} W_{ih}  +  b_{i} )  \\ 
    o_{t} &= \sigma(x_{t} W_{ox}  + h_{t-1} W_{oh}  +  b_{o} ) \\
    f_{t} &= \sigma(x_{t} W_{fx}  + h_{t-1} W_{fh}  +  b_{f} ), 
\end{align}
where $\sigma$ is the sigmoid function\footnote{$\sigma(x) = \frac{1}{1+\exp(-x)}$} applied element-wise, 
the bias terms $b_{;}$ are of size $B\times H$, and $W_{\dot{},x }$ and $W_{\dot{},h}$ are weight matrices with dimensions $I \times H$ and $H \times H$, respectively. From the input, output and forget terms, ($i_t, o_t, f_t$ - all of size $B \times H$), the next instances of the hidden and memory states are computed as:
\begin{align}
  \label{eq_input2} 
    \hat{h}_{t} &= x_{t} W_{hx}  + h_{t-1}  W_{hh} + b_{\hat{h}_{t}} \\
    c_{t} & = f_{t} \odot c_{t-1} + i_{t} \odot \tanh(\hat{h}_{t}) \label{eq_interal_state} \\
    h_{t} &= \tanh(c_{t}) \odot o_{t},	\label{eq_hidden_state}
\end{align}
where $\odot$ denotes element-wise multiplication. The input gate controls which elements of the internal state we are going to update, while the forget gate determines which elements of the internal state ($c_{t}$) will be ``eliminated''.  
In the case of deep networks with multiple LSTM layers, the hidden states generated by each layer are used as inputs to the next layer described by another instance of Eqs.~(1)-(6), where the input terms
$x_{t}$ in Eqs.~(1)-(4) are replaced with the corresponding hidden states of the previous layer.
%explain architacture

\begin{figure}[!ht]
\centering
\includegraphics[width=0.8\textwidth]{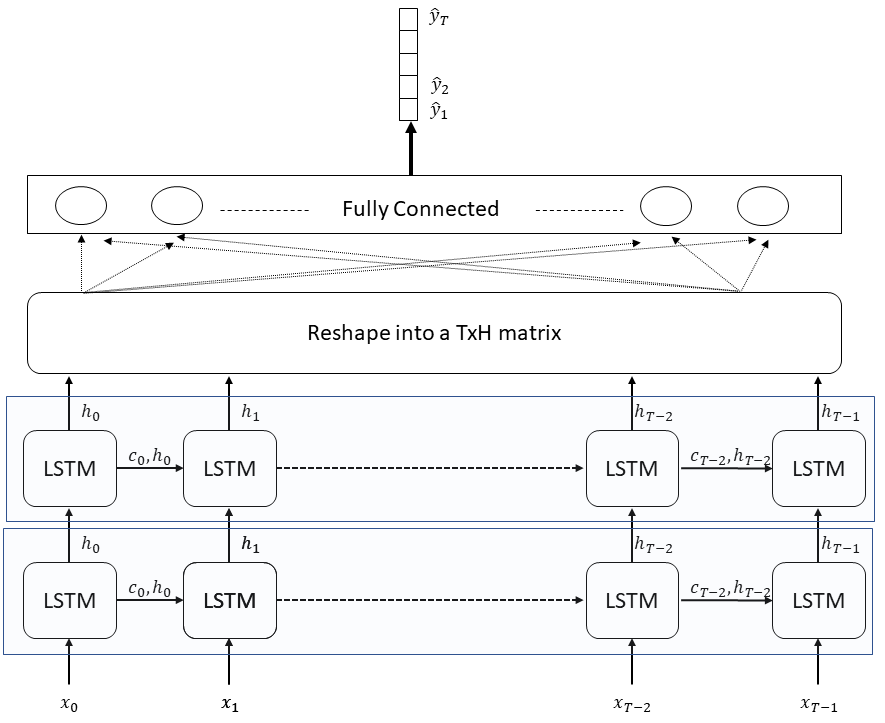}
\caption{High level overview of the ``un-rolled'' LSTM architecture (for the case of two LSTM layers) over length-$T$ input sequences. The hidden states of the last LSTM layer are collected, reshaped and fed through a fully connected layer to produce the desired outputs $\hat{y}$.}
\label{fig:LSTM_Hidden} 
\end{figure}

The un-rolled architecture of our network is depicted in Fig.~\ref{fig:LSTM_Hidden}.
%
%explain operation of the network and design choices
%
Operationally, the network receives a sequence of numerical vectors (or batches thereof) $x_{t-T+1},...,x_{t}$, where $T$ is a \
fixed ``time window'' size (for our purposes $T$ will be measured in trading days), and produces a scalar output sequence of $\hat y_{t-T+2},...,\hat y_{t+1}$ which in our context will represent the predicted price of a stock or other asset. 
We note that for each time $t$, the quantity of interest in the output sequence is its last element, $\hat y_{t+1}$ (i.e., the predicted price for the ``next'' time period), while the previous elements, $\hat y_{t-T+2},...,\hat y_{t}$, correspond to ``predictions'' for the prices $y_{t-T+2},..., y_{t}$, which are already known at time $t$. Although the inclusion of these output elements may seem superfluous, opting for sequence-to-sequence training will make for better predictive accuracy as well as profitability, as we will see further on.
The input vectors are fed into 
an LSTM cell sequentially, resulting in the sequence of hidden state vectors $h_{t-T+1},...,h_{t}$. These are in turn used as inputs to the next LSTM layer. The hidden state vectors generated by the last LSTM layer (Fig.~\ref{fig:LSTM_Hidden} depicts only two LSTM layers although one can introduce additional ones)
are  
fed through a linear fully-connected layer
%(whose weight matrix is of size $H$ $\times$ $O$) 
to produce the output sequence. 
%of size $B \cdot T \times O$. 
%Finally, we reshape the output back to $B$ $\times$ $T$ $\times$ $O$.
In short, we have 
\begin{equation}
{\bf \hat y} = {\bf h} W_{\hat y {\bf h}} + b_{\bf \hat y},
\label{eq:dense}
\end{equation}
where 
${\bf \hat y}$ is $(B \cdot T) \times O$, and O is the output dimension. The  term
${\bf \hat y}$ contains the predicted output vectors $[\hat y_{t-T+2},...,\hat y_{t+1}]^T$, stacked in size-$B$ batches for each time within our 
size-$T$ window,
$\bf h$ is the similarly-stacked $(B \cdot T) \times H$ matrix of the hidden state vectors $h_{t-T+1},...,h_{t}$ produced by the last LSTM layer (each batch contributing $T$ rows), and
$W_{\hat y {\bf h}}$, $b_{\bf \hat y}$ are the weight matrix and bias vectors, sized $H \times O$ and $ (B \cdot T) \times O$, respectively.
For our purposes, the feature vectors, $x_t$, will include an asset's daily adjusted close price, $y_{t}$, opening price, $y^{(O)}_{t}$, intra-day low, $y^{(L)}_{t}$, intra-day high, $y^{(H)}_{t}$, closing price, $y^{(C)}_{t}$, and the previous day's adjusted close, $y_{t-1}$, all obtained online  \citep{yahoofinance},
i.e.,
$$
x_t=[y_{t}, y^{(O)}_{t}, y^{(L)}_{t}, y^{(H)}_{t}, y^{(C)}_{t}, y_{t-1}].
$$
The particular choices of the remaining network parameters (e.g., number of hidden states, $H$, and window size, $T$), will be discussed in Sec.~\ref{hypertuning}.

There are two design choices that we would like to highlight with respect to the network architecture. One has to do with the use of the 
entire sequence of hidden states, $h_{t-T+1},...,h_{t}$, produced by the last LSTM layer, when it comes to computing outputs. More 
typically, one could simply let the hidden state evolve while the input sequence $x_{t-T+1},...,x_t$ is ``fed in'', and use only its 
``last'' value, $h_{t}$, to compute the prediction $\hat y_{t+1}$. Doing so, however, essentially implies an expectation that all of the 
``useful'' information present in the sequence will be encoded into the last hidden state. On the other hand, utilizing the entire 
hidden state sequence will allow us to look for information in the time-evolution of the hidden state within the rolling window, which 
will be prove to be beneficial, as we will see shortly.  
A second important design choice is the handling of the hidden state vectors by reshaping them into a $(B \cdot T) \times H$ matrix, to 
be fed into a single fully-connected linear activation layer as per Eq.~\ref{eq:dense}. By doing so, we avoid having to use one fully-connected %linear activation 
layer per time step, $t-T+1,...,t$, which would introduce $T$ ``copies'' of Eq.~\ref{eq:dense} for handling each of the $h_{t-T+1},...,h_{t}$ separately, resulting in
a total of $T \times  (H \times O + B \times O)$ weights to be adjusted. 
Using a single dense layer as outlined above requires only $H \times O$  + $B \times T \times O$ weights, a significantly lower number as 
the window size, $T$, grows. This choice will allow us to train with a very small batch size, %even only $B=1$, 
which will directly lead to short training times while exploiting information in the time history of the hidden states, and will also make our approach applicable in multi-asset settings or when frequent training is required e.g.,  in intraday trading.

%%%%%%%%%%%%%%%%%%%%%%%%%%%%%%%%%%%%%%%%%%%%

\subsection{Trading Strategy and Allocation} 
\label{trading_strategy_section}

As we have mentioned in Sec.~\ref{sec:litreview}, it is common to make trading decisions based on a prediction model's directional accuracy, even when the model was trained to attain low mean squared or absolute errors. That is, the decision to buy or sell is determined by whether the model predicts an upwards or downwards movement of the asset price over the next time interval.  
Here, we depart from prior approaches by taking advantage of the fact that it is possible to be nearly ``agnostic'' about an event occurring, i.e., the direction (up/down) of an asset's price movement, and yet to be able to glean significant information about the profitability of a trade when conditioning on particular events. In our case the events will have to do with the relative location of predictions within their own distribution.

To make matters precise, we begin by defining the notions of {\em allocation policy} and {\em trading strategy}:
\begin{definition}[Allocation policy]
\label{policy}
Let ${\cal M}$ be a prediction model that produces at each time $t$ an estimate, $\hat y_{t+1}$, of an asset's future value $y_{t+1}$, the latter resulting from some underlying stochastic process. 
Let $\hat r_t=\hat y_{t+1}/y_t-1$ be the model's one-step {\em predicted} returns at time $t$, and $D_{\hat r}$ their distribution. 
An $n$-bin {\it allocation policy} is a pair $(Q, A)$, where $Q \in \mathbb{R}^{n-1}$ is a vector of percentiles of $D_{\hat r}$ in increasing order, and $A \in \{\mathbb{R} \cup {\emptyset}\}^n$ a vector of {\em allocations} in units of the asset (e.g., number of shares to purchase), where the element ${\emptyset}$ denotes sale of all units of the asset held. 
\end{definition}
\begin{definition}[Trading strategy]
\label{trading_strategy}
Given a prediction model, ${\cal M}$, the current asset price, $y_t$, and an $n$-bin allocation policy $(Q,A)$, our proposed trading strategy consists of the following steps, executed at each time $t$:
\begin{itemize}
\item Query ${\cal M}$ for the asset's predicted return $\hat r_t$ in the next time step, $t+1$. 
\item Let
$$
i = \left\{ \begin{array}{lll} 1 & \mbox{if} & \hat r_t < Q_1 \\
j+1 & \mbox{if} & \hat r_t \in [Q_j, Q_{j+1}), \\ 
n & \mbox{if} & \hat r_t \ge Q_{n-1}.
\end{array} \right. 
$$
\item If $A_i>0$ and we are currently not holding any of the asset, buy $A_i$ units of the asset. If $A_i>0$ and we are already holding some of the asset, or if $A_i=0$, do nothing. 
\item If $A_i = \emptyset$, sell any and all units of the asset held.
\end{itemize}
\end{definition}
In our case, the model ${\cal M}$ will be the LSTM network described in the previous Section. Intuitively, the percentiles in $Q$ partition the real line into $n$ ``bins'', and the trading strategy is to buy $A_i$ units of the asset whenever the predicted return lies in bin $i$ and we do not already own the asset. For simplicity, we will consider allocation policies in which $Q_1=0$ and $A_1= \emptyset, A_{i>1} \ge 0$, i.e, we always sell all of our asset if the predicted return for the next trading day is negative, that being the only event on which we sell. Of course, one can envision variations of the above strategy, including negative allocation values corresponding to ``short sales'' (i.e., selling an amount of the asset under the obligation to buy it back at a later time, with the expectation that its value will have declined), assigning the ``sell signal'', $\emptyset$, to more than one bins as determined by $Q$, using time-varying values of $A_i$ depending on the available cash in our portfolio, or measuring the allocations $A_i$ in monetary amounts as opposed to units of the asset. Some of these options are worth exploring in their own right, but we will limit ourselves to the setting previously described because of space considerations. 

{\em Remark:} We note that the proposed trading policy can be viewed as a generalization of the classical directional ``up-down'' trading strategy. If we set $Q = 0$, i.e., we use only one separating point, and $A=[\emptyset, 1]^T$, then there will be only two bins, one for positive predicted returns (in which case we buy 1 unit of the asset), and one for negative predicted returns (in which case we sell), which is exactly how the directional strategy works.

\subsubsection{Optimizing the allocation policy}
\label{sec:optimal_allocation}
Armed with the above definitions, we can now consider the question of what allocation policy is optimal for a given prediction model while following the trading strategy of Def.~\ref{trading_strategy}. We will be buying some units of the asset, $A_i$, when our model's predicted return lies in bin $i>1$, and will then hold those units until the predicted return falls within bin $1$. 
Over time, this process generates a set of {\em holding intervals}, $I_{ij}=[\tau_{ij}+1, \omega_{ij}]$, where $\tau_{ij}$ denotes the time of  the $j$-th buy transaction while the predicted return was in bin $i>1$, and $\omega_{ij}$ the time of the following sell transaction (i.e., the first time following $\tau_{ij}$ that the predicted return falls within bin 1). It should be clear from Def.~\ref{trading_strategy} that the intervals $I_{ij}$ will be non-overlapping (because we never execute a buy when already holding the asset) and that their union will be the entire set of times during which we hold a nonzero amount of the asset.

Our allocation policy risks a monetary amount of $A_i \cdot y_{\tau_{ij}}$  to purchase $A_i$ units of the asset at each time $\tau_{ij}$. 
%(when the predicted return `'`landed'' in bin $i>1$ for the $j$-th time, and we were holding zero stock). 
The net profit generated from the at-risk amount at the end of the holding interval $I_{ij}$ (i.e., after a single buy-sell cycle) is therefore
\begin{equation}
A_i (y_{\omega_{ij}}-y_{\tau_{ij}}).
\end{equation}
Over time, there will be a number of instances, $N_i$, that the asset was purchased while at bin $i$
and the total net profit after $M=\sum N_i$ buy-sell transactions will then be
\begin{equation}
G=\sum_{i=2}^n A_i \sum_{j=1}^{N_i} (y_{\omega_{ij}}-y_{\tau_{ij}}),
\label{eq:G}
\end{equation}

resulting in an expected net profit of
\begin{equation}
{\cal E}\{G\} = \sum_{i=2}^n A_i {\cal E} \left\{ \sum_{j=1}^{N_i} (y_{\omega_{ij}}-y_{\tau_{ij}}) \right\}.
\label{eq:expG}
\end{equation}

The expectation terms in Eq.~\ref{eq:expG} represent the expected sum of differences in asset price over each bin's holding intervals, that is, the expected total rise (or fall) in asset price conditioned on buying when the expected return falls within each particular bin and following the proposed trading strategy.  
It is clear from Eq.~\ref{eq:expG} that the expected total profit will be maximized with respect to the allocations $A_i$ if each $A_i$ is set to zero whenever the associated expectation term is negative, and to as large a value as possible when the expected price difference is positive.   Assuming an practical upper limit, $A_{max}$, to the number of asset units we are able or willing to buy each time, the optimal allocation values are thus 
\begin{equation}
 A_{i>1} = \left\{ \begin{array}{ll}
A_{max} & \mbox{if} \quad {\cal E} \left\{ \sum_{j=1}^{N_i} (y_{\omega_{ij}}-y_{\tau_{ij}}) \right\} >0,\\
 0 & \mbox{otherwise.} 
\end{array}
\right.
 \label{eq:sdcond}
\end{equation}
In practice, the expectations terms in Eq.~\ref{eq:sdcond} will  be approximated empirically from in-sample data, therefore their sign could be wrongly estimated, especially when they happen to lie close to zero.
For this reason, one may wish to slightly alter Eq.~\ref{eq:sdcond} so that we buy $A_{max}$ when the expected sum of price differences is greater than some small threshold $\epsilon > 0$, to reduce the chance of ``betting on loosing bins''. 
Finally, we note that we are taking $A_{max}$ to be constant (across time and across the bins $i$) for the sake of simplicity, although other choices are possible as we have previously mentioned. 

\section{Model training and allocation policy selection}
\label{sec:training}

As we have seen, the specification of the optimal allocation and trading strategy described in the previous Section, requires the (empirical) distribution of the predicted returns which are generated by our LSTM model. That is, we need a trained model in order to obtain concrete values for the percentiles in  $Q$ and the ``bins'' they induce, as well as the estimated sums of price differences per bin that will determine the allocations, $A$, in Eq.~\ref{eq:sdcond}. 
A step-by-step summary of our approach is as follows: 
\begin{itemize}
\item Train and test the LSTM network over a rolling window of size $T$, where for each trading day, $t$, in the period from 1/1/2005-1/1/2008, the network's weights are first adjusted using price data from the immediate past, (i.e., input $[x_{t-T},\ldots, x_{t-1}]$, and output target $[y_{t-T+1},\ldots,y_t]$); 
then, the network is given $[x_{t-T+1},\ldots, x_{t}]$ as input and asked to predict the price sequence once step ahead, 
$[\hat y_{t-T+2},\ldots,\hat y_{t+1}]$, from which we keep the 
next day's adjusted closing price, $\hat y_{t+1}$, and calculate the predicted return, $\hat r_t$.
\item Use the distribution $D_{\hat r}$ of the predicted returns generated by the rolling-window process to determine the vector of cutoff points, $Q$. These points could be chosen to have constant values, or to correspond to specific percentiles of $D_{\hat r}$, or in any other consistent manner.
\item For each bin determined by $Q$, use the history of input data and corresponding model predictions over 1/1/2005-1/1/2008 to determine the buy and sell events that would have resulted by having followed the trading strategy of Def.~\ref{trading_strategy}. 
Compute the asset price differences for each buy-sell cycle and sum over each bin, $i$, in which the corresponding buy transactions took place (Eq.~\ref{eq:expG}). Set the allocations $A_{i>1}$ to their optimal values as per Eq.~\ref{eq:sdcond}.
\item Given the allocation policy, $(Q,A)$, test the profitability of the trading strategy and all variants of our LSTM model with respect to a range of hyper-parameters, over a subsequent time period, $1/2/2008$-$12/31/2009$, again training the network daily based on input data from the previous $T$ trading days before predicting the next day's price. Select the hyper-parameter values which yield the most profitable -- as opposed to most accurate -- model.
\item Apply the selected model to a new data sample (1/4/2010-5/1/2018) and evaluate its profitability\footnote{Profitability will simply be $G/C$, where $G$ is the net profit, as calculated in (\ref{eq:G}) and $C$ is the initial capital invested}. We will use the term {\em out-of-sample} (instead of {\em testing}) when referring to this period, in order to avoid confusion with the rolling training-testing procedure used to make predictions. Similarly, the period 1/1/2005-12/31/2009 used to construct the initial allocation policy and perform hyper-parameter selection will be referred to as {\em in-sample}.
\end{itemize}
We go on to discuss important details and practical considerations for each of these steps.

\subsection{LSTM training}
In contrast to other studies, e.g., \citep{Bao}, which make use of a more traditional training - validation - testing framework, we found that a rolling training-testing window approach was better suited to our setting, where at each time, $t$, the network is trained using data from the immediate past and then asked to predict ``tomorrow's''  price (at time $t+1$). By updating the model weights at each step, using the most recent information before making a prediction, we avoid ``lag'' or ``time disconnection'' between the training and testing data (as would be the case, for example if our training data was ``far'' into the past compared to the time in which we are asked to make a prediction) and allow the model to better ``react'' to changing regimes in the price sequence as time progresses, leading to it being more profitable as we shall see. 
As we noted in Sec.~\ref{network_arch}, we used a sequence-to-sequence training schema because it led to significantly higher cumulative returns compared to sequence-to-value training. We speculate that  
asking the network to predict {\em sequences} of the asset price over a time interval, combined with the use of all hidden vectors produced in that interval, allows the network to learn possible latent \say{micro-structure} within the price sequence (e.g., trend or other short-term characteristics),  thus facilitating the training process.

The network was trained using back-propagation through time (BPTT), with the ADAM optimizer, implemented in python 3 and Tensorflow v1.8.  Network weights were initialized from a uniform distribution (using the Glorot-Xavier uniform method) and LSTM cell states were initialized to zero values. 
We applied exponential decay to our learning rate, and trained for $1600$ iterations. The number of iterations was chosen to avoid over-fitting by observing both the training and testing errors, using a mean-squared loss function.
It is important to note that the rolling window training process, as described above, used a  batch size of $1$ (i.e., the network was trained each time on a {\em single} $T$-sized window of the data, $x_t$). Our numerical experiments showed that for our proposed architecture, larger batch sizes did not increase the ultimate profitability of the model and trading strategy, and sometimes even led to lower performance\footnote{For example, training with a batch size of $50$ on the S\&P$500$ index resulted in twice the mean average percent error and a $15\%$ lower cumulative return compared to training with a batch size of 1, with full details on trading performance to be given in Section~\ref{sec:results}.}. Depending on the choice hyperparameters (listed in Section~\ref{hypertuning}), the time required to compute one rolling window step was between $7$ and $30$ seconds on an modestly equipped computer (Intel i7 CPU and $16$ GB of RAM).

\subsection{Allocation policy selection} \label{allocation_policy_selection}

The overall number and values of the allocation policy's cutoff points ($Q$) were determined after numerical experimentation to ensure that all of the bins defined by $Q$ included an adequate population of samples and were sufficient in number to allow us to discern differences in the profits realized by the trading strategy when executing a buy in different bins. We found that an effective approach, in terms of ultimate profitability, was to select the cutoff points in $Q$ using percentiles of the distribution of the absolute expected returns, $D_{|\hat r|}$ %(effectively increasing the number of data points per bin). 
Thus, the $Q_1$ cutoff was set to zero (so that as per Def.~\ref{trading_strategy} we sell any holdings if our model's predicted return is negative), and another six cutoff points, $Q_2,\ldots,Q_7$, were chosen to correspond to the first six deciles ($10\%$, $20\%$, $30\%$, $40\%$, $50\%$, and $60\%$ points) of the distribution of absolute predicted returns, splitting the set of predicted returns into eight bins. For each bin, $i$, we could then use historical predicted and actual prices to calculate the sum of price differences (as per in Eq.~\ref{eq:sdcond}) and determine the optimal allocation $A_i$.
We emphasize that the above choices with respect to $Q$ were the result of experimentation, and that the elements in $Q$ could in principle be optimized; however, doing so is not trivial, and we will not pursue it here because of space considerations.

In practice, it is be advantageous to allow the allocation policy $(Q,A)$ to adapt to new data rather than be fixed for a significant amount of time, because the distributions of the predicted and realized returns are unlikely to be stationary.  As the model makes a new prediction each day, we have a chance to add the new data to the distribution of predicted returns and adjust the cutoff points in $Q$. In the same vein, as the trading strategy generates new buy-sell decisions, we may re-calculate the allocations $A$. There are many options here, such as to accumulate the new information on returns and trading decisions as in a growing window that stretches to the earliest available data, or to again use a rolling window so that as each day new data is added to the distribution, ``old'' data is removed. 

With the above in mind, we chose to update the percentiles in $Q$ at every time step, using a growing window of data that began 120 time steps prior to our first out-of-sample prediction. For example, in order to making a prediction for $1/4/2010$ - the first day of our out-of-sample period -  the model's predicted returns between $7/15/2009$ - $1/1/2010$ were included in the empirical distribution used to calculate $Q$. 
%As a new predicted return was generated at each time $t$, it was included in the data, and the $Q_{i>1}$ could be re-adjusted. 
This was done in order to avoid the problem of initially having no data (e.g., on the first day of our out-of-sample period) from which to calculate the $Q_i$. 
The 120 step ``bootstrap'' period was selected after experimenting with several alternatives, including using more or less of the initial history of predicted returns, or using a fixed-size rolling window to determine how much of the ``past'' should be used. 
With respect to the allocations, $A_i$, when computing the per-bin sums of price differences in Eq.~\ref{eq:sdcond}, we found that it was best in practice (i.e.,  increased profitability) to include all of the available history of buy-sell events and corresponding price differences, up to the beginning of our data sample ($1/1/2005$). Moving forward, as buy and sell decisions were made at each time step, the resulting price differences were included in the sums of Eq.~\ref{eq:sdcond} and the allocations $A$ were updated, if necessary.
An instance of the set of bins, corresponding cutoff points and sums of price differences for each bin (computed for use on the first trading day of the out-of-sample period, $1/04/2010$) are listed in Table~\ref{cutoff_bins_allocation}. 
For example, if the predicted return for $1/4/2010$ for the S$\&$P$500$ falls within the  $30$-$40\%$ bin (or, equivalently in the interval $[0.65\%, 0.97\%)$),
we do not buy any of the asset (corresponding allocation set to zero) because if we did, the expected price difference in the asset would be negative at the time we sell the asset. 
\begin{table}[!ht]
\centering
\begin{tabular}{l | c c | c c | c c | c c } 
     & \multicolumn{2}{c}{S$\&$P$500$} & \multicolumn{2}{c}{DJIA} & \multicolumn{2}{c}{NASDAQ} & \multicolumn{2}{c}{R$2000$} \\ \hline
Bin $i$ (deciles) & $Q_i$ & $\delta_i$  & $Q_i$  & $\delta_i$ & $Q_i$  & $\delta_i$ & $Q_i$ & $\delta_i$\\ \hline \hline
\text{1: SELL} & $0$ & -  & $0$ & -  & $0$ & -  & $0$ & - \\
2: $0\%$-$10\%$  & $0.12\%$ &  $126.97$ & $0.01\%$ & $-1785.45$  & $0.05\%$ & $62.37$ & $0.01\%$ & $139.19$ \\
3: $10\%$-$20\%$ & $0.38\%$ &  $99.92$ & $0.02\%$ & $-1529.37$  & $0.10\%$ & $520.54$ & $0.03\%$ & $90.76$\\
4: $20\%$-$30\%$ & $0.65\%$ &  $131.35$ & $0.03\%$ & $904.60$   & $0.12\%$ & $52.27$ &  $0.06\%$ & $23.19$\\
5: $30\%$-$40\%$ & $0.97\%$ &  $-66.71$ & $0.04\%$ & $-88.28$   & $0.16\%$ & $-59.84$ & $0.07\%$ & $87.32$\\
6: $40\%$-$50\%$ & $1.18\%$ &  $128.67$ & $0.06\%$ & $1062.01$  & $0.18\%$ & $206.65$ & $0.10\%$ & $1.64$\\
7: $50\%$-$60\%$ & $1.44\%$ & $-191.68$ & $0.08\%$ & $381.46$   & $0.24\%$ & $254.78$ & $0.12\%$ & $6.17$\\
8: $60\%$-$100\%$ & - 	   & $222.85$ &  - 		& $3850.18$       & 	-	  & $719.68$ & 	- 	  & $248.96$\\ \hline \hline
\end{tabular}
\caption{Snapshot of the cut-off points, $Q_i$ (reported as percentages), for each bin-interval of the predicted return distribution. The $Q_i$ mark the upper values of each bin, and are computed using the model's predicted returns from $7/15/2009$ -$1/04/2010$. The $\delta_i$ denote empirical estimates of the corresponding sums of price differences from Eq.~\ref{eq:sdcond} when using the proposed trading strategy, over the  period $1/1/2005$-$1/04/2010$. }
\label{cutoff_bins_allocation}
%\end{adjustwidth}
\end{table}

As previously discussed, it is optimal to buy $A_i=A_{max}$ units 
only when the predicted return falls within a bin whose realized sum
of price differences is positive. In practice, the choice of 
$A_{max}$ is usually subject to budget constraints or investment mandates which limit the maximum number of units to purchase. This maximum could also be tied to the risk profile of the investor.
In our study, $A_{max}$ was set to the maximum number of units one could buy with their available capital at the start of the trading period. For example, at the beginning of our out-of-sample period, 1/04/2010, the price of the S\&P 500 index was $\$1,132.989$, and therefore if our initial capital was $\$28,365$ we would set $A_{max}=25$.
On a practical note, however, because stock indices are not directly tradeable, one can instead trade one of the available Exchange Traded Funds (ETFs) which track the index and are considered one of the most accessible and cheapest options for investing in indices. 
Therefore, in order to apply our approach to the S\&P$500$, we could choose to trade SPY, a popular ETF which tracks that index, and -- based on the aforementioned initial capital and SPY price of $\$94.55$ on 1/04/2010 -- set $A_{max}=300$.
In the experiments detailed in the next Section, we will choose to trade four well-known ETFs, namely SPY for the S\&P$500$, DIA for the DJIA, IWM for the R$2000$ and ONEQ for the NASDAQ. 
%We will discuss specifics in Section~\ref{sec:results}. 
%the underlying index SPY was trading at $\$94.55$, therefore if our initial capital was $\$28,365$ we would set $A_{max}=300$.

\subsection{Hyper Parameter Tuning} \label{hypertuning}
A grid-search approach was used to select the  parameters of our neural network, namely the number of LSTM layers, number of units per LSTM layer ($H$), the dropout parameter (dropout was applied only on the input of a given neuron), and the length $T$ of the input sequence. Because our ultimate goal is profitability, this meant evaluating the cumulative returns of our model under each combination of parameters, in order to identify the most profitable variant (which could be different depending on the stock or stock index under consideration). 
For each choice of hyper-parameters listed in Table~\ref{grid_search_range}, we carried out the rolling training-testing procedure described above, over the period 1/1/2005-1/1/2008, 
and calculated an initial allocation strategy $(Q,A)$. 
\begin{table}[!ht]
%\begin{adjustwidth}{-2.25in}{0in} % Comment out/remove adjustwidth environment if table fits in text column.
\centering
\begin{tabular}{l l l l}
Parameters & Range \\ \hline \hline
Nr. of LSTM layers & $ 2,3$\\
Nr. of units ($H$)  & $32,64,128$\\
Input sequence length ($T$)  & $11,22,44$\\
Dropout & $0$,$50$,$70\%$\  \\  \hline \hline
\end{tabular}
\caption{Hyper-parameter grid-search values.}
\label{grid_search_range}
%\end{adjustwidth}
\end{table}
Following that, we used data from  1/2/2008-12/31/2009, to calculate the profit realized by applying our trading strategy (while the allocation policy evolved each trading day, as described in the previous Section). 
This subsequent period, containing $505$ trading days, contained a sufficient number of rolling windows and 
corresponding predictions from which to compare  performance for each 
choice of hyper-parameters. In addition, it encompasses  the global financial crisis of 2009 and is long enough to capture various market cycles.
%By splitting the two in-sample periods around January of $2008$ we hoped to influence the training of the model sufficiently so that it captured the financial crisis without being overwhelmed by it. 
The model parameters leading to the highest cumulative return for each of the four stock indices of interest were thus identified and are listed in Table~\ref{tb_hyper_tuning}. 
\begin{table}[!ht]
%\begin{adjustwidth}{-2.25in}{0in} % Comment out/remove adjustwidth environment if table fits in text column.
\centering
\begin{tabular}{l c c c c}
Parameters & S\&P 500 & DJIA & NASDAQ & R$2000$\\ \hline \hline
Nr. of LSTM layers    & $3$  & $3$ & $3$ & $3$\\ 
Nr. of units ($H$)  & $64$ & $64$ & $32$ & $32$\\ 
Input sequence length ($T$)  & $22$ & $22$ & $22$ & $11$\\
Dropout         & $50\%$ & $70\%$ & $50\%$  & $50\%$ \\  \hline \hline
\end{tabular}
\caption{Grid-search: optimal hyper-parameter values.}
\label{tb_hyper_tuning}
%\end{adjustwidth}
\end{table}
On a small 6-machine, 96-CPU, 96Gb RAM computing cluster, 
the entire grid search process was easily parallelized. For smaller network configurations, e.g., input sequences of length $11$, using $2$ layers and $32$ neurons, the process described above took a few ($\le 5$) hours to complete, while configurations with length-$44$ input sequences required up to $3$ days.

\section{Results}
\label{sec:results}

Having settled on a choice of hyper-parameters, we tested the performance (annualized returns and other measures of interest) of the resulting model(s) and associated trading strategy on our out-of-sample period, $1/4/2010$-$5/1/2018$, which we had set aside for this purpose. 
Summarizing the procedure outlined in the previous Section, for each trading day, $t$, within this period, we updated the LSTM network's weights by training it on the time window of input data from $t-T,...,t-1$ with a target output of $y_{t}$, and then computed the prediction, $\hat y_{t+1}$, of the next day's closing price, using input data over the window $t-T+1$,...,$t$.
Next, we determined the predicted return, $\hat r_t=\hat y_{t+1}/y_t-1$, and the bin, $i$, (based on the percentiles in $Q$) within which it lies. We then  bought or sold the amount of asset units (or corresponding ETF in case of an index) prescribed by the allocation policy (i.e., buy $A_i$ units  if $i>1$ and we did not already own the asset, and sell all of our units if $i=1$) 
If a buy was executed, we recorded the index of the bin, $i$, and the asset price at that time. When that asset was later sold, the profit obtained from that transaction (i.e., sell price minus buying price) was
added to the sum of price differences term in Eq.~\ref{eq:sdcond}, and we recalculated the corresponding allocation $A_i$ if the update caused a change in the sign of the sum.
Finally, we included $\hat r_t$ to the growing set of previous predicted returns and adjusted the decile points in $Q$ to account for their new distribution, thus completing our update of the allocation policy, $(Q,A)$.

We proceed with the main results concerning i) the prediction accuracy, and ii) the profitability of our proposed scheme over the out-of-sample period, for each of the stock indices studied. We will also discuss comparisons against recent efforts, as well as ``naive'' prediction approaches and trading strategies. 

\subsection{Error metrics and prediction performance} 
\label{results_error_metrics}
Although our central goal is to achieve good trading performance, we will nevertheless begin by evaluating our proposed LSTM model using some of the metrics often cited in studies involving prediction accuracy 
including mean directional accuracy\footnote{$\mathrm{MDA} = 
\dfrac{1}{N} \sum_{t=1}^{N} \mathds{1}(sign(\hat{y_{t}}-y_{t-1}) \cdot sign(y_{t}-
y_{t-1}))$, where the indicator function $\mathds{1}(x)$ returns $1$ if $x$ is 
positive, zero otherwise, $y_{t}$ and $\hat{y_{t}}$ are the realized and predicted prices for time $t$, respectively, $N$ is the length of the out-the-sample sequence.} (MDA) which in less formal terms measures how accurate a model is at predicting the correct direction (up/down) of stock price movement,
%regardless of the magnitude of the error. whether - on average - the predicted and 
%realized values of the sequence $y_t$ ``point in the same direction'' relative to 
%the current value., 
mean squared error\footnote{$\mathrm{MSE} = 
\dfrac{1}{N} \sum_{t=1}^{N} (y_{t} - \hat{y_{t}})^{2}$.} (MSE), mean absolute 
error\footnote{$\mathrm{MAE} = \dfrac{1}{N} \sum_{t=1}^{N} | y_{t} - 
\hat{y_{t}}|$.} (MAE), mean absolute percentage Error\footnote{$\mathrm{MAPE} = 
\dfrac{1}{N} \sum_{t=1}^{N} \frac{| y_{t} - \hat{y_{t}}|}{y_{t}}$.} (MAPE) and 
correlation coefficient\footnote{$R^2 = \dfrac{\sum_{t=1}^{N} (y_{t} - \bar{y}) 
(\hat{y_{t}}-\bar{\hat{y}})}{ \sqrt{\sum_{t=1}^{N} (y_{t} - \bar{y})^{2} } \sqrt{\sum_{t=1}^{N} (\hat{y_{t}} - \bar{\hat{y}} )^{2}}} 
$, where $\bar{y}$ and $\bar{\hat{y}}$ are the averages of the actual ($y_t$) and predicted ($\hat y_t$) price sequences, respectively.} ($R^2$). Table~\ref{Error_metrics} shows the predictive performance of 
the proposed LSTM model for each of the four stock indices, for the same out-of-
sample period ($1/4/2010$-$5/1/2018$).
% \begin{table}[!ht]
% \centering
% \begin{tabular}{l | c c| c c | c c | c c}
%  & \multicolumn{2}{c}{ S$\&$P $500$} & \multicolumn{2}{c}{DJIA} & \multicolumn{2}{c}{NASDAQ} & \multicolumn{2}{c}{R2000} \\ \hline \hline
% Metrics 	 & $ARIMA_{2,1,1}$    &  LSTM   &   $ARIMA_{3,2,1}$ 	& LSTM	 & $ARIMA_{3,2,2}$	  & LSTM 		& $ARIMA_{2,1,0}$ & LSTM\\ \hline
% MDA  & $50\%$  & $50.8\%$ & $50.31\%$  & $50.6\%$    &  $51.04\%$ & $50.2\%$ 	& $50.36\%$ & $49.77\%$\\
% MAPE & $0.7\%$ & $1.33\%$  & $0.65\%$   & $0.6\%$   &  $1.07\%$  & $0.75\%$  	& $0.9\%$  & $0.8\%$\\ 
% MAE  & $22.51$ & $42.92$  & $105.56$   & $96.79$ 	 &  $43.97$   & $31.57$  	& $9.21$  & $9.06$\\ 
% MSE  & $1101$ 	& $4993$  & $24245$      & $19645$   &  $4048$	  & $1982$ 		& $148.7$ & $143.56$ \\ 
% $R^{2}$ & $99.94\%$ & $99.72\%$ & $99.91\%$ &$99.94\%$ & $99.89\%$ & $99.95\%$  & $99.88\%$  & $99.9\%$\\ \hline \hline
% \end{tabular}
% \caption{Error metrics for the LSTM and ARIMA models. The indices on the ARIMA models (p,d,q) denote the number of AR(p), difference (d) and MA(q) terms, respectively.}
% \label{Error_metrics}
% \end{table}
\begin{table}[!ht]
\centering
\begin{tabular}{c c c c c}
Metrics 	 &  S$\&$P $500$   &   DJIA	 	& NASDAQ 		& R$2000$\\ \hline \hline
MDA   &   $51.05\%$ & $51\%$  &  $50.47\%$ &  $50.02\%$ \\
MAPE &   $0.66\%$  & $0.6\%$ & $0.8\%$    & $0.9\%$ \\
MAE  &   $11.07$ & $95.33$  & $30.8$      & $9.01$ \\
MSE  &   $256$   &  $19047$ & $1893$      & $142$ \\
$R^{2}$ &  $99.94\%$ &  $99.94\%$   &  $99.95\%$  & $99.90\%$     \\ \hline \hline
\end{tabular}
\caption{Error metrics for the LSTM model.}
\label{Error_metrics}
\end{table}
Overall, our model achieves a MAPE below $1\%$ and a MDA narrowly above $50\%$, for all indices. However, applying the Pesaran-Timmermann (PT) to the predicted vs. actual direction of price movements did not confirm that the LSTM model accurately predicts price movement direction in a statistically significant manner. The test statistics and corresponding p-values were  (PT=-3.61, p=0.999), (PT=-1.84, p=0.9674), (PT=-0.64, p=0.7378), and (PT=-2.03, p=0.9789), for the S$\&$P$500$, DIJA, NASDAQ and R2000 indices, respectively.
%\textcolor{blue}{The latter implies that our model may provide only a narrow advantage when predicting price movement direction}. 
The fact that our LSTM model does not provide a significant advantage in predicting next-day price movement direction implies that our model may not be effective when used with the day-to-day directional \say{up-down} strategy often used in the literature. It does not, however, preclude us from being able to produce significant profits under {\em other} trading strategies, such as the one proposed in Sec.~\ref{trading_strategy_section} which does not require directional accuracy on the part of the model.

When compared against other recent studies, e.g., \citet{Bao}, our model has significantly lower MAPE for S$\&$P$500$ and DIJA (see Table~\ref{comparison_errormetrics_bao}), while the $R^{2}$ coefficient is better in both cases (using the same test period as in \citet{Bao}). 
\begin{table}[!ht]
\centering
\begin{tabular}{c c c c c}
 & \multicolumn{2}{c}{S$\&$P$500$}  &   \multicolumn{2}{c}{DIJA} \\ \hline	\hline
Metics   & \citet{Bao} & LSTM & \citet{Bao} & LSTM  \\
MAPE   & $1.1\%$ & $0.7\%$ & $1.1\%$  & $0.6\%$ \\
$R^{2}$ & 0.946 &  $0.999$ & 0.949 & $0.999\%$ \\ \hline	\hline
\end{tabular}
\caption{Comparison with \citet{Bao} (S$\&$P$500$ and DIJA) for the out-of-sample period used therein ($10/1/2010$ - $9/30/2016$).}
\label{comparison_errormetrics_bao}
\end{table}
Looking at \citet{EMD2FNN} and the metrics reported therein (see Table~\ref{comparison_errormetrics_EMD2FNN}), 
\begin{table}[!ht]
\centering
\begin{tabular}{c c c c c}
 & \multicolumn{2}{c}{S$\&$P$500$}  &   \multicolumn{2}{c}{NASDAQ} \\ \hline	\hline
Metrics & \citet{EMD2FNN} & LSTM & \citet{EMD2FNN} & LSTM  \\
MAPE   & $1.05\%$  & $1.07\%$ & $1.08\%$ & $1.17\%$  \\
MAE    & $13.03$   & $13.19$  & $52.4773$ & $30.8463$  \\
RMSE   & $17.6591$ & $18.4715$ & $70.4576$ & $41.1703$ \\ \hline	\hline
\end{tabular}
\caption{Comparison with \citet{EMD2FNN} (S$\&$P$500$ and NASDAQ), for the out-of-sample period used therein (the year $2011$).}
\label{comparison_errormetrics_EMD2FNN}
\end{table}
for the same out-of-sample period used in that work, our model 
performs comes in very close for the S$\&$P$500$ and performs better for NASDAQ with respect to MAE and RMSE. 
Finally, Table~\ref{comparison_errormetrics_ModAugNet} compares 
with \citet{ModAugNet}, where our model outperforms in all three statistics (MAPE, MAE, MSE).
\begin{table}[!ht]
\centering
\begin{tabular}{c c c }
 & \multicolumn{2}{c}{S$\&$P$500$}   \\ \hline	\hline
Metrics & \citet{ModAugNet} & LSTM \\
MAPE   &  $1.0759\%$  & $0.8\%$ \\
MAE    &  $12.058$   & $11.65$  \\
MSE    &  $342.48$ & $276.27$ \\ \hline	\hline
\end{tabular}
\caption{Comparison with \citet{ModAugNet} (for the S$\&$P$500$),  for the out-of-sample period used therein $01.02.2008$ - $7.26.2017$.}
\label{comparison_errormetrics_ModAugNet}
\end{table}
%\textcolor{red}{Despite making less accurate predictions, however, our model will perform better in terms of profitability, as we shall see shortly}.
We note that our model relies on a relatively simple architecture with no pre-processing layers, making it simple to implement and fast to train. This is to be contrasted with more elaborate designs, such as the three-layered approach of \citet{Bao} passing data through a wavelet transformation and autoencoder neural network before reaching the LSTM model, or the ``dual'' LSTM models of \cite{ModAugNet}. 
Despite its simplicity, our model's predictive performance is similar to that of more complex approaches, and compares favorably to studies with long out-of-sample periods (\cite{Bao}, \cite{ModAugNet}).
%, e.g., lower MAPE  for indices relative to \cite{Bao} and \cite{ModAugNet}, respectively.

We emphasize the fact that the comparisons and discussion of the model's predictive performance - although encouraging - are given here mainly for the sake of providing a fuller picture, and that good predictive performance will mean little unless our model also performs favorably in terms of profitability (to be discussed shortly).
%We emphasize the fact that the Despite accurate predictions, however, our model should be evaluated in terms of profitability, as we shall see shortly present.}
In fact, there is an important point of caution to keep in mind regarding 
predictive accuracy, and it has to do with the proper 
context when citing MAPE, MAE and MSE values. Specifically, when 
it comes to asset prices, daily changes are 
usually small on a relative basis, and thus a MAPE of less 
than $1\%$, for example, is neither unusual nor surprising. 
In fact, a naive model which ``predicts'' that tomorrow's price will simply be equal 
to today's (i.e., $\hat{y}_{t+1} = y_{t}$), applied to the 
S$\&$P$500$, would typically yield very low error metrics, e.g., MAPE = $0.64\%$, MAE=$20.0203$, MSE=$844.186$ over our out-of-sample period, or a MAPE of $0.83\%$ for the period examined in \citet{ModAugNet} (to be compared with a MAPE of $1.0759\%$ reported in that work).
Of course, this naive model is completely useless from the point of view of trading because it does not allow us to make any decision on whether to buy or sell, highlighting the fact that models with similar prediction accuracy can have vastly different profitability, while favorable MAPE, MAE, and MSE scores do not by themselves  guarantee success in trading. When the ultimate goal is profitability, the prediction model should be evaluated in the context of a specific trading strategy, and one may achieve high(er) returns by looking beyond average errors, as the trading and allocation strategies of Section~\ref{trading_strategy_section} do.

\subsection{Profitability} \label{results_profitability}

\begin{figure}[!ht]
\centering
\includegraphics[width=5.5in]{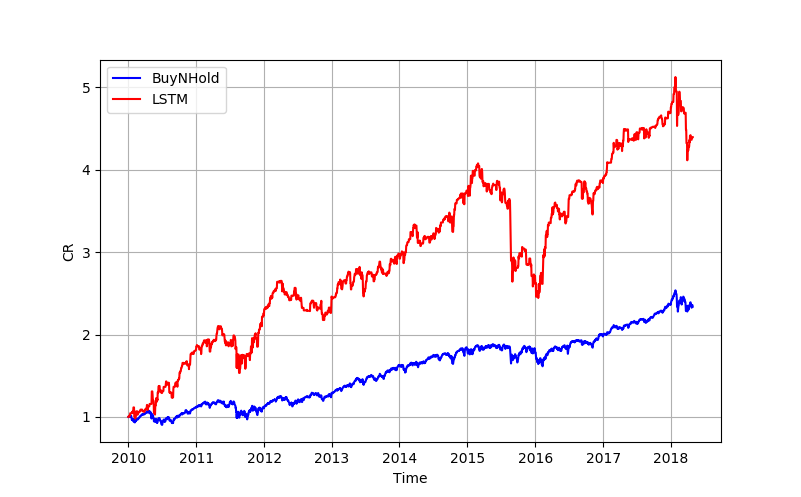}
\caption{Growth of one monetary unit invested using our proposed model (LSTM) vs. the buy-and-hold strategy, on the S$\&$P$500$.}
\label{capital_growth_sp500}
\end{figure}

We now turn our attention to the profitability of our proposed LSTM model and trading strategy.  The results presented below are labeled according to the stock indices studied, however -- as noted in Sec.~\ref{allocation_policy_selection} -- all trading is done using their corresponding tracking ETFs, namely SPY for the S\&P500, DIA for the DJIA, IWM for the R2000 and ONEQ for the NASDAQ, since the indices themselves are not tradeable.
The cumulative returns we obtained when trading over 1/04/2010-5/1/2018 (listed in Table~\ref{SummaryResults}), 
\begin{table}[!ht]
\centering
\begin{tabular}{l l c c c c c }
   & & \multicolumn{5}{c}{Profitability Measures}   \\ \hline \hline
   & & CR & AR & AV & SR & DD\\ \hline 
S$\&$P $500$ &  LSTM 		  & $339.6\%$ &	$19.5\%$ &	$24.3\%$ &	$0.8$ &	$-20.0\%$  \\
			  &  BnH   		  & $136.4\%$ &	$10.9\%$ &	$14.9\%$ &	$0.7$ &	$-11.2\%$\  \\
              \hline       
DJIA		  &  LSTM 		  & $185.1\%$ &	$13.3\%$ &	$24.6\%$ &	$0.5$ &	$-14.3\%$ \\ 		 
			  &  BnH   		  & $136.6\%$ &	$10.8\%$ &	$14.0\%$ &	$0.8$ &	$-10.7\%$\ \\
             \hline       
NASDAQ 		 &  LSTM 		  & $370.9\%$ &	$20.3\%$ &	$38.3\%$ &	$0.5$ &	$-28.0\%$   \\ 
			 &  BnH   		  & $228.9\%$ &	$15.2\%$ &	$16.8\%$ &	$0.9$ &	$-12.5\%$  \ \\
               \hline       
R$2000$		 &  LSTM 		  & $360.6\%$ &	$20.0\%$ &	$33.0\%$ &	$0.6$ &	$-24.8\%$   \\
			 &  BnH   		  & $163.5\%$ &	$12.2\%$ &	$20.0\%$ &	$0.6$ &	$-15.8\%$\ \\
                 \hline \hline      
\end{tabular}
\caption{Returns and risk statistics for the out-of-sample period (1/04/2010-5/1/2018). \emph{CR} denotes cumulative return, \emph{AR} is annualized return, \emph{AV} is annualized volatility, \emph{SR} is the Sharpe ratio, \emph{DD} is draw-down, defined as the largest peak-to-trough percent decline during an investment's lifetime; 
\emph{LSTM} is the our proposed model and trading strategy, while \emph{BnH} represents the buy-and-hold strategy in which the asset is bought once at the beginning of the trading period and sold at the end. The S\&P500 was traded via the SPY ETF, the DJIA via DIA, the R2000 via IWM, and the NASDAQ via ONEQ.}
\label{SummaryResults}
\end{table} 
significantly outperformed the benchmark buy-and-hold strategy  
($340\%$ vs $136\%$, $185\%$ vs $137\%$, $371\%$ vs $229\%$, and $361\%$ vs $164\%$ for the S\&P500, DJIA, NASDAQ and R2000 indices, respectively); the same was true on an annualized basis. A sample comparison illustrating the time history of capital growth using our approach vs buy-and-hold is shown in Figure~\ref{capital_growth_sp500}.

Our superior performance over the buy and hold strategy, comes at the cost of a higher standard deviation of returns achieved 
($24.3\%$ vs $14.9\%$, $24.6\%$ vs $14\%$, $38.3\%$ vs $16.8\%$, $33\%$ vs $20\%$ for the S$\&$P$500$, DJIA, NASDAQ and R$2000$, respectively), leading to lower corresponding Sharpe ratios (except for the S$\&$P$500$) compared to buy-and-hold.
Draw-down, a measure of investment risk defined as the largest peak-to-trough decline during the investment's life-cycle, was also affected, with our model having larger negative peak-to-trough returns versus  buy-and-hold ($-20\%$ vs $-11.2\%$, $-14.3\%$ vs $-10.7\%$, $-28\%$ vs $-12.5\%$, $-24.3\%$ vs $-15.8\%$ for the S\&P5$00$, DJIA, NASDAQ and R2000, respectively). This is expected, given that we are comparing a semi-active trading strategy against the ``passive'' buy-and-hold strategy that executes only a single buy; however, the draw-down is not excessive, and our trading strategy is able to more than make up for it as is evident from its superior annualized and cumulative returns. Finally, our trading 
strategy executed $535$, $394$, $494$, and $544$ trades for the S\&P$500$, DJI, NASDAQ, and R$2000$, respectively, over the $2117$ trading days in the out-of-sample period. On average, this corresponds to approximately one transaction per week (ranging from once per four trading days for the R$2000$, to once per six days for the DJI). We note that the returns cited above ignore transaction costs. We have chosen to do this because our strategy trades only one asset,
%(i.e. corresponding ETF), 
does not take short positions where costs have a larger impact, and trades sparingly as noted above.
Moreover, transaction costs for institution-class investors are minimal, ranging from $2$ to $5$ basis points per trade regardless of the transaction amount, while retail investors can nowadays trade without cost through online brokers\footnote{Examples of brokers that offer free retail trades include RobinHood (https://robinhood.com), and  Vanguard (https://investor.vanguard.com/investing/transaction-fees-commissions/etfs)}. 

\subsubsection{Comparisons with recent works}
\label{sec:compare_recent}
Besides the benchmark buy-and-hold strategy, the results given in the previous Section indicate that our proposed model outperforms those in several recent works which attempt to beat the major stock indices used here, most often the S\&P500. %\textcolor{blue}{Although the prior approaches cited here differ in methodology and underlying asset(s) traded, 
%(ranging from trading a portfolio of stocks to indices) 
%they all focus on beating an index, most often the S\&P$500$.}
%(i.e, not individual stocks or portfolios of stocks). 
A direct comparison to \citet{ModAugNet} (also trading the S$\&$P$500$ over a long out-of-sample period, $1/2/2008$-$7/26/2017$), shows that our approach outperforms by a significant margin in terms of cumulative return ($334.36\%$ vs $96.55\%$).
Next to \citet{OmerMurat_Algorithmic}, 
which 
employed deep convolutional neural networks instead of LSTMs, 
our model  achieved an annualized return of $18.31\%$ 
compared to the $10.77\%$ reported in that work when trading the same tracking EFT as we have (SPY), or $13.01\%$ when using a portfolio of ETFs, over the period of $1/1/2007$-$12/31/2016$. Other studies with relatively long out-of-sample periods, include \citet{krauss_do_huck}  and \citet{KraussFisher}; compared to those, our model yields a significantly greater cumulative return ($204\%$ vs $-17.96\%$ and $-15\%$, respectively, over their out-of-sample period, $2010$-$2015$) and, as a result, a higher positive Sharpe ratio ($0.8$ vs negative for the other two works), noting however that those works traded a portfolio of stocks derived from the S\&P500 and not the index (or ETF) itself.
Finally, we note that our results are not directly comparable with those in works who report high {\em average} or {\em summed} returns (e.g., \citet{Bao} and others)
from which unfortunately one cannot deduce annualized returns or other ``standard'' measures of trading performance, as we have explained in Section~\ref{sec:litreview}.

The works cited thus far reported performance over multi-year periods. Direct comparisons between different approaches over short evaluation periods must be made cautiously because superior performance over any short period does not necessarily imply sustainable long-term results, making it difficult to say which method is better. Having said that however, compared to the model from \citet{EMD2FNN} for the S$\&$P$500$, using the same out-of-sample period (the year $2011$), our approach attained a higher cumulative return of $33.98\%$ vs $25\%$. Finally, in comparison with \citet{adaptive_stock_index} during their out of sample period (the year $2010$) our approach also performs strongly ($80.90\%$ vs. $41.89\%$ when using their long-only trading strategy). 
%However, shifting to $2011$ (our best performing year for the S\&P500), our model achieves an annual return of $52.44\%$ which was far higher than our long-term annualized return and illustrates our point about the role of the size of evaluation period. 

\subsubsection{Attributing the overall performance}

% As we have stated, altering either the trading strategy or  the prediction model will affect the overall scheme's profitability. This brings up the question of ``how much''  of the performance attained is due to the proposed LSTM architecture versus the trading and allocation strategy which we have outlined. This question is difficult to answer fully, in part because it requires rating the performance of many alternative prediction models when paired with our proposed trading strategy, as well as different trading strategies to be used with our proposed LSTM model. Instead, we will opt to gain insight by examining the performance of the linear (ARIMA) predictor models model(s) discussed in Sec.~\ref{results_error_metrics} under our proposed trading strategy, as well as the performance of our LSTM network under the classic directional ``up-down''strategy used frequently  (with minor modifications) throughout the literature, \citep{Bao,ModAugNet,EMD2FNN}.

As we have previously stated, altering either the trading strategy or  the prediction model will affect the overall scheme's profitability. This brings up the question of ``how much''  of the performance attained is due to the proposed LSTM architecture versus the trading and allocation strategy which we have outlined. This is difficult to answer fully, partly because it requires rating the performance of many alternative prediction models when paired with our proposed trading strategy, as well as different trading strategies to be used with our proposed LSTM model. Here, we will opt to gain some insight by examining the performance of a simple autoregressive integrated moving average (ARIMA) predictor model under our proposed trading strategy, as well as the performance of our LSTM network under the classic directional ``up-down''strategy whose variants are used frequently throughout the literature %with minor modifications, 
\citep{Bao,ModAugNet,EMD2FNN}. 
Of course the ARIMA model lacks sophistication, but its purpose here will be to serve as a ``canonical'' baseline case only.

To proceed with our analysis, we fitted an $ARIMA_{p,d,q}$ model to the daily (adjusted close) price sequence of each stock index under consideration, where $p$, $d$ and $q$ are the orders of the autoregressive, difference, and moving average terms, respectively.   
In each case, the model was fitted to the price data of the in-sample period ($1/1/2005$ - $12/31/2009$), and the optimal values of the $p$, $d$, and $q$, were determined by searching over the integers (up to a maximum of order 3) and selecting the model with the lowest AIC and BIC. The selected models were then examined to ensure that their coefficients were statistically significant at the $5\%$ level. 
The resulting model orders $(p,d,q)$ were: $(2,1,1)$ for the S$\&$P$500$, $(3,2,1)$ for the DJIA, $(3,2,2)$ for the NASDAQ, and $(2,1,0)$ for the R$2000$ index. The model coefficients are not 
listed here for the sake of brevity. 

In terms of MAPE, MAE and MSE, the predictive performance of the ARIMA models was close to but generally slightly worse than that of our proposed LSTM model (see Table~\ref{ARIMA_Error_metrics}, where for convenience we have also inlcuded the LSTM-based error metrics from Table~\ref{Error_metrics}).
\begin{table}[!ht]
\centering
\begin{tabular}{l | c c| c c | c c | c c}
 & \multicolumn{2}{c}{ S$\&$P $500$} & \multicolumn{2}{c}{DJIA} & \multicolumn{2}{c}{NASDAQ} & \multicolumn{2}{c}{R2000} \\ \hline \hline
Metrics 	 & $ARIMA_{2,1,1}$    &  LSTM   &   $ARIMA_{3,2,1}$ 	& LSTM	 & $ARIMA_{3,2,2}$	  & LSTM 		& $ARIMA_{2,1,0}$ & LSTM\\ \hline
MDA          &  $49.69\%$ & $51.05\%$ & $50.50\%$  & $51\%$     &  $50.88\%$   & $50.47\%$   & $49.88\%$ & $50.02\%$\\
MAPE         & $0.7\%$    & $0.66\%$  &  $0.6\%$   & $0.6\%$     &  $1.08\%$   & $0.8\%$  	 & $0.9\%$   & $0.9\%$\\ 
MAE          & $22.81$    & $11.07$   & $104.04$   & $95.33$     &  $42,28$    & $30.8$  	 & $9.17$    & $9.01$\\ 
MSE          & $1983.18$  & $256$     & $23555$    &  $19047$    &  $3798$     & $1893$ 	 & $147.6$   & $142$ \\ 
$R^{2}$      & $99.94\%$  & $99.94\%$ &    $99.91\%$ & $99.94\%$   & $99.89\%$ & $99.95\%$   & $99.88\%$        & $99.9\%$\\ \hline \hline
\end{tabular}
\caption{Error metrics for the LSTM and ARIMA models. The indices on the ARIMA models (p,d,q) denote the number of AR(p), difference (d) and MA(q) terms, respectively. LSTM error metrics are from Table \ref{Error_metrics}}
\label{ARIMA_Error_metrics}
\end{table}
%\textcolor{blue}{To evaluate whether the LSTM forecasts are more or less accurate than those of the ARIMA model(s) (alternative hypothesis) we present the Diebold Mariano test statistic along with the corresponding p-values at $5\%$ confidence level: S$\&$P$500$ (DM =$4.30$, p-value=$0.999$), DJI (DM =$-34.58$, p-value=$0.000$), IXIC (DM=$-26.32$, p-value=$0.000$), R$2000$ (DM=$2.86$, p-value=$0.999$). Hence, for  S$\&$P$500$ and R$2000$,  we cannot reject the null hypothesis and LSTM predictions are more accurate than ARIMA's model while for the other two the opposite holds.}
A Diebold-Mariano (DM) test was performed to determine whether the LSTM price forecasts were more or less accurate than those of the ARIMA model(s). Under the null hypothesis that the ARIMA forecast was more accurate than that of the LSTM-based model, the test statistics and corresponding p-values for the S$\&$P$500$, DJIA, NASDAQ and R2000 were  (DM =$-4.30$, p-value=$0.000$), (DM =$34.58$, p-value=$1.000$), (DM=$26.32$, p-value=$1.000$), and (DM=$-2.86$, p-value=$0.002$), respectively. This indicates that at $5\%$ confidence level the LSTM model was a better (one-step) predictor than the ARIMA model only in the cases of the S$\&$P$500$ and the R$2000$ but of course, 
as we have alluded to earlier, profits -- rather than statistical measures of accuracy -- should be used to evaluate stock market forecasts \citep{Leith_Tanner_Forecast}.

Turning to the central issue of profitability, we can elucidate the effect of the proposed trading strategy alone, by examining the performance ``boost'' it gives the LSTM and ARIMA models, compared to when those same models are paired with the directional \say{up-down} trading strategy where each trading day we buy if the predicted price is greater than the current price, and sell if it is lower. The profitability (cumulative returns) of each model-strategy combination is shown in Table~\ref{ARIMASummaryResults}, for each of the four stock indices. \begin{table}[!ht]
    \centering
    \begin{tabular}{l l c c }
      &  \multicolumn{3}{l}{~~~~~~~~~~~~~~~~~~~~~~~~~~~~Trading Strategies}   \\ \hline \hline
      & & Proposed & \say{Up-Down} \\ \hline 
    S$\&$P $500$  &  LSTM 		  & $339.6\%$ &	$83.56\%$ \\
                  &  $ARIMA_{2,1,1}$ & $194.5\%$ &	$58.15\%$  \\ 
    			  &  BnH: $136.4\%$    &  & \ \\
                  \hline       
    DJIA		  &  LSTM 		  & $185.1\%$ &	$85.39\%$ \\ 		 
    			  &  $ARIMA_{3,2,1}$ & $51.65\%$ &	$16.49\%$ \\
    			  &  BnH: $136.6\%$      &  & \ \\ 
                 \hline       
    NASDAQ 		 &  LSTM 		  & $370.9\%$ &	$115.2\%$ \\ 
    			 &  $ARIMA_{3,2,2}$  & $293.03\%$ & $83.35\%$   \\
                &  BnH: $228.9\%$            &  & \ \\
                  \hline       
    R$2000$		 &  LSTM 		  & $360.6\%$ &	$111.4\%$   \\
    			 &  $ARIMA_{2,1,0}$   	  & $42.39\%$ & $37.96\%$  \\
                 &  BnH: $163.5\%$           &  & \ \\
                     \hline \hline      
    \end{tabular}
    \caption{Cumulative returns over the out-of-sample period (1/04/2010-5/1/2018) using the proposed trading and allocation policy versus the directional ``Up-Down'' strategy. LSTM stands for our proposed model while BnH represents the buy-and-hold strategy in which the asset is bought once at the beginning of the trading period and sold at the end.}
    \label{ARIMASummaryResults}
\end{table} 
Unsurprisingly, the linear ARIMA model was well behind the LSTM model in terms of cumulative returns; its returns also had roughly twice the annualized standard deviation\footnote{under the proposed trading strategy, AV for ARIMA vs LSTM was $ 44.78\%$ vs $24.3\%$, $50.69\%$ vs $24.6\%$,  $76.84\%$ vs $38.3\%$, and $60.91\%$ vs $33\%$,  for the S\&P$500$, DJIA, NASDAQ and R$200$, respectively.}.
Two things are worth observing, however. 
First, our proposed trading and asset allocation policy leads to a significant boost in cumulative returns with {\em both} models, sometimes allowing even the (simplistic) ARIMA model to outperform the buy-and-hold strategy (S\&P$500$, NASDAQ). 
Second, when using the ``up-down'' strategy, the LSTM model significantly outperforms the ARIMA model for all stock indices, despite the fact that the two models were similar in terms of accuracy, as previously discussed; 
in that case, however, the LSTM-based returns were much weaker relative to the buy-and-hold approach (or indeed to the studies mentioned in Section~\ref{sec:compare_recent}).
%as shown in Table \ref{ARIMASummaryResults}. 
The fact that the LSTM model can perform strongly (Section~\ref{results_profitability}) but not when used 
with a strategy other than the one proposed here, highlights the fact that it is fruitful to treat the 
design of the prediction model and trading strategy as a joint problem, despite the difficulty involved.
%These results, together with those of Table~\ref{SummaryResults} 
%indicate that i) our proposed trading and asset allocation policy 
%offer a significant boost in profitability for both models (LSTM and ARIMA) and ii) the LSTM model does not ``beat'' the buy-and-hold strategy (or indeed previous models) when used with a strategy other than the one proposed here, highlighting the fact that it is fruitful to treat the design of the prediction model and trading strategy as a joint problem, despite the difficulty involved.

% Specifically, we have already seen (Table~\ref{SummaryResults} above) that the use of using our LSTM network instead of the linear (ARIMA) model, when they {\em both} use our proposed strategy, yields an improvement in cumulative returns by a factor of approximately between 2 and 4, depending on the stock index. 

\section{Conclusions and future work}
\label{sec:conclusions}
%say that it's the combination of model+strategy that is important 
%Opening: say what we did

Motivated by the complexity involved in designing effective  models for stock price prediction together with accompanying trading strategies, as well as the prevalence of ``directional'' approaches, 
%Our work was motivated by the complexity of the neural network models,  how they are evaluated in training and testing phases and the prevalence of directional strategies.
we presented a simple LSTM-based model for predicting asset prices, together with a strategy that takes advantage of the model's predictions in order to make profitable trades.  Our approach is not focused solely on constructing a more precise or more directionally accurate (in terms of whether the price will rise or fall) model; instead, we exploit the distribution of the model's predicted returns, and the fact that the ``location'' of a prediction within that distribution carries information about the expected profitability of a trade to be executed based on that prediction. The trading policy we described %generalizes, and hence 
departs from the oft-used directional \say{up/down} strategy, allowing us to harness more of the information contained within the model's predictions, even then that model is relatively simple.

%Novelty of model/strategy
Our proposed model architecture consists of a simple yet effective deep LSTM neural network, which uses a small number of features related to an asset's price over a relatively small number of time steps in the past (ranging from 11 to 22), in order to predict the price in the next time period. 
Novel aspects of our approach include: i) the use of the entire history of the LSTM cell's hidden states as inputs to the output layer, as the network is exposed to an input sequence, ii) a trading strategy that takes advantage of useful information that becomes available on the profitability of a trade once we condition on the predicted return's position within its own distribution, and iii) selecting the network's hyper-parameters to identify the most profitable model variant (in conjunction with the trading strategy used), instead of the one with lowest prediction error,
recognizing the fact that predictive accuracy alone does not guarantee profitability. 
Our design choices with respect to the LSTM network allowed us to ``economize'' on the architecture of the prediction model, using 2-3 LSTM layers, input sequences of 11-22 time-steps, and 32-128 neurons, depending on the stock index studied. 
%Furthermore, we were able to train on  size-$1$ batches, making 
Our model is frugal in terms of training data and fast in terms of training-testing times (updating the model and trading strategy requires between $7$ and $30$ seconds on a typical desktop computer), and is thus also suitable for intra-day or multi-stock/index prediction applications.

The performance of our proposed model and trading strategy was tested on four major US stock indices, namely the S\&P500, the DJIA, the NASDAQ and the Russel 2000 over the period $10/1/2010$-$5/1/2018$. To the best of our knowledge, besides ``beating'' the indices themselves, our model and trading strategy outperform those proposed in several recent works using multi-year testing periods, in terms of the cumulative or annualized returns attained while also doing well in terms of volatility and draw-down. 
With respect to works that used short (e.g. single-year) testing periods, our scheme outperformed some in their chosen comparison periods, or could achieve lower vs. higher returns if the time period was shifted, keeping in mind that it is difficult to reach safe conclusions when comparing over any one short time interval. Based on the overall profitability results and rather long testing period used in this work, we feel that our approach shows significant promise.

Opportunities for future work include further experimentation with more sophisticated allocation and trading strategies, the possible inclusion of short sales, as well as the use of a variable portion of the time history of the LSTM hidden states when making predictions, to see where the optimum lies. Also, it would be of interest to optimize the manner in which the percentiles of the predicted return distribution are chosen when forming our allocation strategy, in order to maximize profitability and reduce risk. Finally one could in principle use the proposed distribution-based trading strategy with any other model that outputs price predictions, and it would be interesting to see what performance gains, if any, could be achieved.

%\appendix

\section*{References}

%% References


%\newpage
%\begin{thebibliography}{10}
%\input{bib.tex}

\bibliographystyle{apa}
\bibliography{refs.bib}

\begin{thebibliography}{}

\bibitem[\protect\astroncite{Baek and Kim}{2018}]{ModAugNet}
Baek, Y. and Kim, H.~Y. (2018).
\newblock Mod{A}ug{N}et: A new forecasting framework for stock market index
  value with an overfitting prevention {LSTM} module and a prediction {LSTM}
  module.
\newblock {\em Expert Systems with Applications}, 113:457--480.

\bibitem[\protect\astroncite{Bao et~al.}{2017}]{Bao}
Bao, W., Yue, J., and Rao, Y. (2017).
\newblock A deep learning framework for financial time series using stacked
  autoencoders and long-short term memory.
\newblock {\em PLoS ONE}, 12(7): e0180944.

\bibitem[\protect\astroncite{Chiang et~al.}{2016}]{adaptive_stock_index}
Chiang, W.-C., Enke, D., Wu, T., and Wang, R. (2016).
\newblock An adaptive stock index trading decision support system.
\newblock {\em Expert Systems with Applications}, 59(C):195--207.

\bibitem[\protect\astroncite{Chong et~al.}{2017}]{chong_han_park_error_metrics}
Chong, E., Han, C., and Park, F.~C. (2017).
\newblock Deep learning networks for stock market analysis and prediction:
  methodology, data representations, and case studies.
\newblock {\em Expert Systems with Applications}, 83(C):187--205.

\bibitem[\protect\astroncite{Deng et~al.}{2017}]{deep_RL}
Deng, Y., Bao, F., Kong, Y., Ren, Z., and Dai, Q. (2017).
\newblock Deep direct reinforcement learning for financial signal
  representation and trading.
\newblock {\em IEEE Transactions on Neural Networks and Learning Systems},
  28(3):653--664.

\bibitem[\protect\astroncite{Ding et~al.}{2015}]{event-driven}
Ding, X., Zhang, Y., Liu, T., and Duan, J. (2015).
\newblock Deep learning for event-driven stock prediction.
\newblock In {\em Proceedings of the 24th International Conference on
  Artificial Intelligence}, IJCAI'15, pages 2327--2333. AAAI Press.

\bibitem[\protect\astroncite{Fama and French}{1993}]{Fama_French}
Fama, E.~F. and French, K.~R. (1993).
\newblock Common risk factors in the returns on stocks and bonds.
\newblock {\em Journal of Financial Economics}, 33(1):3--56.

\bibitem[\protect\astroncite{Fama and French}{2004}]{fama_french_CAPM}
Fama, E.~F. and French, K.~R. (2004).
\newblock The capital asset pricing model: Theory and evidence.
\newblock {\em Journal of Economic Perspectives}, 18(3):25--46.

\bibitem[\protect\astroncite{Fischer and Krauss}{2018}]{KraussFisher}
Fischer, T. and Krauss, C. (2018).
\newblock Deep learning with long short-term memory networks for financial
  market predictions.
\newblock {\em European Journal of Operational Research}, 270(2):654--669.

\bibitem[\protect\astroncite{Foster et~al.}{2018}]{nmt_recent_advances}
Foster, G., Vaswani, A., Uszkoreit, J., Macherey, W., Kaiser, L., Firat, O.,
  Jones, L., Shazeer, N., Wu, Y., Bapna, A., Johnson, M., Schuster, M., Chen,
  Z., Hughes, M., Parmar, N., and Chen, M.~X. (2018).
\newblock The best of both worlds: Combining recent advances in neural machine
  translation.
\newblock In {\em Proceedings of the 56th Annual Meeting of the Association for
  Computational Linguistics, {ACL} 2018, Melbourne, Australia, July 15-20,
  2018, Volume 1: Long Papers}, pages 76--86.

\bibitem[\protect\astroncite{Gu et~al.}{2018}]{gu_Kelly_Xiu}
Gu, S., Kelly, B.~T., and Xiu, D. (2018).
\newblock Empirical asset pricing via machine learning.
\newblock Paper in Swiss Finance Institute Research Paper No. 18-71.

\bibitem[\protect\astroncite{Hochreiter and Schmidhuber}{1997}]{LSTM_origin}
Hochreiter, S. and Schmidhuber, J. (1997).
\newblock Long short-term memory.
\newblock {\em Neural computation}, 9(8):1735--1780.

\bibitem[\protect\astroncite{Huck}{2009}]{Huck_2009}
Huck, N. (2009).
\newblock {Pairs selection and outranking: An application to the S\&P 100
  index}.
\newblock {\em European Journal of Operational Research}, 196(2):819--825.

\bibitem[\protect\astroncite{Huck}{2010}]{Huck_2010}
Huck, N. (2010).
\newblock {Pairs trading and outranking: The multi-step-ahead forecasting
  case}.
\newblock {\em European Journal of Operational Research}, 207(3):1702--1716.

\bibitem[\protect\astroncite{Krauss et~al.}{2017}]{krauss_do_huck}
Krauss, C., Do, X.~A., and Huck, N. (2017).
\newblock {Deep neural networks, gradient-boosted trees, random forests:
  Statistical arbitrage on the S\&P 500}.
\newblock {\em European Journal of Operational Research}, 259(2):689--702.

\bibitem[\protect\astroncite{Leitch and Tanner}{1991}]{Leith_Tanner_Forecast}
Leitch, G. and Tanner, J.~E. (1991).
\newblock Economic forecast evaluation: Profits versus the conventional error
  measures.
\newblock {\em American Economic Review}, 81(3):580--90.

\bibitem[\protect\astroncite{Minh et~al.}{2018}]{dl_GRU}
Minh, D.~L., Sadeghi-Niaraki, A., Huy, H.~D., Min, K., and Moon, H. (2018).
\newblock Deep learning approach for short-term stock trends prediction based
  on two-stream gated recurrent unit network.
\newblock {\em IEEE Access}, 6:55392--55404.

\bibitem[\protect\astroncite{Olah}{2018}]{colahs}
Olah, C. (2018).
\newblock {Understanding LSTM Networks, August 17 (2015)}.
\newblock
  \url{http://colah.github.io/posts/2015-08-Understanding-LSTMs/}[Accessed: 1st
  May 2018].

\bibitem[\protect\astroncite{Rather et~al.}{2015}]{RNN_and_hybrid_model}
Rather, A.~M., Agarwal, A., and Sastry, V. (2015).
\newblock Recurrent neural network and a hybrid model for prediction of stock
  returns.
\newblock {\em Expert Systems with Applications}, 42(6):3234--3241.

\bibitem[\protect\astroncite{Russakovsky
  et~al.}{2015}]{image_recongnition_ILSVRC}
Russakovsky, O., Deng, J., Su, H., Krause, J., Satheesh, S., Ma, S., Huang, Z.,
  Karpathy, A., Khosla, A., Bernstein, M., Berg, A.~C., and Fei-Fei, L. (2015).
\newblock {ImageNet Large Scale Visual Recognition Challenge}.
\newblock {\em International Journal of Computer Vision (IJCV)},
  115(3):211--252.

\bibitem[\protect\astroncite{Sermpinis et~al.}{2013}]{Sermpinis_EJOR_SwarmOpt}
Sermpinis, G., Theofilatos, K., Karathanasopoulos, A., Georgopoulos, E.~F., and
  Dunis, C. (2013).
\newblock {Forecasting foreign exchange rates with adaptive neural networks
  using radial-basis functions and Particle Swarm Optimization}.
\newblock {\em European Journal of Operational Research}, 225(3):528--540.

\bibitem[\protect\astroncite{Sethi et~al.}{2014}]{Sethi_Treleaven_Rollin}
Sethi, M., Treleaven, P., and Rollin, S. D.~B. (2014).
\newblock Beating the {S}\&{P} 500 index — a successful neural network
  approach.
\newblock In {\em 2014 International Joint Conference on Neural Networks
  (IJCNN)}, pages 3074--3077.

\bibitem[\protect\astroncite{Sezer and Ozbayoglu}{2018}]{OmerMurat_Algorithmic}
Sezer, O. and Ozbayoglu, M. (2018).
\newblock Algorithmic financial trading with deep convolutional neural
  networks: Time series to image conversion approach.
\newblock {\em Applied Soft Computing}, 70:525--538.

\bibitem[\protect\astroncite{Silver et~al.}{2017}]{AlphaGo}
Silver, D., Schrittwieser, J., Simonyan, K., Antonoglou, I., Huang, A., Guez,
  A., Hubert, T., Baker, L., Lai, M., Bolton, A., Chen, Y., Lillicrap, T., Hui,
  F., Sifre, L., van~den Driessche, G., Graepel, T., and Hassabis, D. (2017).
\newblock Mastering the game of {G}o without human knowledge.
\newblock {\em Nature}, 550:354–359.

\bibitem[\protect\astroncite{Wu et~al.}{2016}]{NMT}
Wu, Y., Schuster, M., Chen, Z., Le, Q.~V., Norouzi, M., Macherey, W., Krikun,
  M., Cao, Y., Gao, Q., Macherey, K., Klingner, J., Shah, A., Johnson, M., Liu,
  X., Kaiser, L., Gouws, S., Kato, Y., Kudo, T., Kazawa, H., Stevens, K.,
  Kurian, G., Patil, N., Wang, W., Young, C., Smith, J., Riesa, J., Rudnick,
  A., Vinyals, O., Corrado, G., Hughes, M., and Dean, J. (2016).
\newblock Google's neural machine translation system: Bridging the gap between
  human and machine translation.
\newblock {\em CoRR}, abs/1609.08144.

\bibitem[\protect\astroncite{YahooFinance}{2018}]{yahoofinance}
YahooFinance (2018).
\newblock {Yahoo Finance, Symbol Lookup (2018)}.
\newblock \url{https://finance.yahoo.com/quote/}[Accessed: 1st May 2018].

\bibitem[\protect\astroncite{Zhong and Enke}{2017}]{Zhong_Enke}
Zhong, X. and Enke, D. (2017).
\newblock Forecasting daily stock market return using dimensionality reduction.
\newblock {\em Expert Systems with Applications}, 67:126--139.

\bibitem[\protect\astroncite{Zhou et~al.}{2019}]{EMD2FNN}
Zhou, F., Zhou, H., Yang, Z., and Yang, L. (2019).
\newblock {EMD2FNN}: A strategy combining empirical mode decomposition and
  factorization machine based neural network for stock market trend prediction.
\newblock {\em Expert Systems with Applications}, 115:136--151.

\end{thebibliography}

%\end{thebibliography}

%% \bibliographystyle{model1-num-names}
%% \bibliography{sample.bib}






% \end{thebibliography}

\end{document}